\documentclass[prl, twocolumn, showpacs, nofootinbib, amsmath,amssymb, floatfix, eqsecnum]{revtex4-1}
\pdfoutput=1
\usepackage{amsmath}
\usepackage{braket}
\usepackage{amssymb}
\usepackage{amsthm}
\usepackage{algpseudocode}
\usepackage{algorithm}
\usepackage{amsfonts}
\usepackage{comment}
\usepackage[normalem]{ulem}
\usepackage{graphicx}
\usepackage{color,framed}
\usepackage{hyperref}
\usepackage{times}
\usepackage{enumerate}
\usepackage{lipsum}
\usepackage{slashed}
\usepackage{xurl}
\usepackage{bbm}
\usepackage{chngcntr}
\counterwithout{equation}{section}
\usepackage{tikz,pgfplots}
\usepackage{pgfplotstable}
\usepgfplotslibrary{fillbetween}

\hypersetup{
    colorlinks=true, 
    linktoc=all,     
    linkcolor=blue,  
}

\def \beq {\begin{equation}}
\def \eeq {\end{equation}}
\def \beqa {\begin{eqnarray}}
\def \eeqa {\end{eqnarray}}
\def \bseq {\begin{subequations}}
\def \eseq {\end{subequations}}

\pgfplotsset{compat=1.18}
\begin{document}

\title{Benchmarking a heuristic Floquet adiabatic algorithm for the Max-Cut problem}

\author{Etienne Granet and Henrik Dreyer}
\affiliation{Quantinuum, Leopoldstrasse 180, 80804 Munich, Germany}
\date{\today}

\begin{abstract}
According to the adiabatic theorem of quantum mechanics, a system initially in the ground state of a Hamiltonian remains in the ground state if one slowly changes the Hamiltonian. This can be used in principle to solve hard problems on quantum computers. Generically, however, implementation of this Hamiltonian dynamics on digital quantum computers requires scaling Trotter step size with system size and simulation time, 
which incurs a large gate count. In this work, we argue that for classical optimization problems, the adiabatic evolution can be performed with a \emph{fixed, finite Trotter step}. This ``Floquet adiabatic evolution" reduces by several orders of magnitude the gate count compared to the usual, continuous-time adiabatic evolution. We give numerical evidence using matrix-product-state simulations that it can optimally solve the \texttt{Max-Cut} problem on $3$-regular graphs in a large number of instances, with surprisingly low runtime, even with bond dimensions as low as $D=2$. Extrapolating our numerical results, we estimate the resources needed for a quantum computer to compete with classical exact or approximate solvers for this specific problem.

\end{abstract}

\maketitle

\textbf{\emph{Introduction.}}--- The adiabatic theorem of quantum mechanics provides an all-purpose solver for classically hard optimization problems \cite{farhi2000quantum}. As convergence to the solution requires to control a slow, continuous-time dynamics which comes with a large gate overhead on digital quantum computers, these approaches are mostly seen as long-term fault-tolerant era methods. In the current and near-term stage of quantum computing hardware, variational algorithms such as QAOA are preferred for their low circuit depth \cite{farhi2014quantum,peruzzo2014variational,blekos2024review,leontica2024exploring}. These can be interpreted as a kind of adiabatic evolution with a finite number of gates, in which Trotter steps are optimized to minimize the energy. A major challenge of variational algorithms for classical optimization problems is that they replace a hard optimization problem by another optimization problem that can be just as difficult to solve~\cite{mcclean2018barren,wang2021noise,lykov2023sampling,ho2016quasi}.

In this work, we present evidence for the fact that \emph{finite} Trotter steps can be used in the adiabatic algorithm (instead of costly continuous-time dynamics) while preserving guarantees to reach the ground state in the slow evolution regime. This holds true only for optimization problems where the target Hamiltonian is classical. The dynamics imposed on the system is then always a Floquet dynamics that alternates between different Hamiltonians during a finite time. Using matrix product state (MPS) simulations of these dynamics, we will show two main points. First, the ``Floquet" adiabatic algorithm with a finite Trotter step $\delta t$ succeeds in finding the optimal solution to a \texttt{Max-Cut} problem on a $3$-regular graph in a very large number of instances, despite the number of samples being a minuscule part of the state space. Extrapolating our numerics, we make an estimate for the resources required by a quantum computer to reach system sizes out of reach of classical solvers, whether exact or approximate. Second, we observe that, surprisingly, using a MPS with a small bond dimension does not (in the vast majority of cases) prevent the adiabatic algorithm to find the optimal solution. This yields a new classical approach to approximately solve the \texttt{Max-Cut} problem with good practical performance that is very different from usual solvers.

\begin{figure}[tb]
\begin{center}
\includegraphics[width=0.245\textwidth]{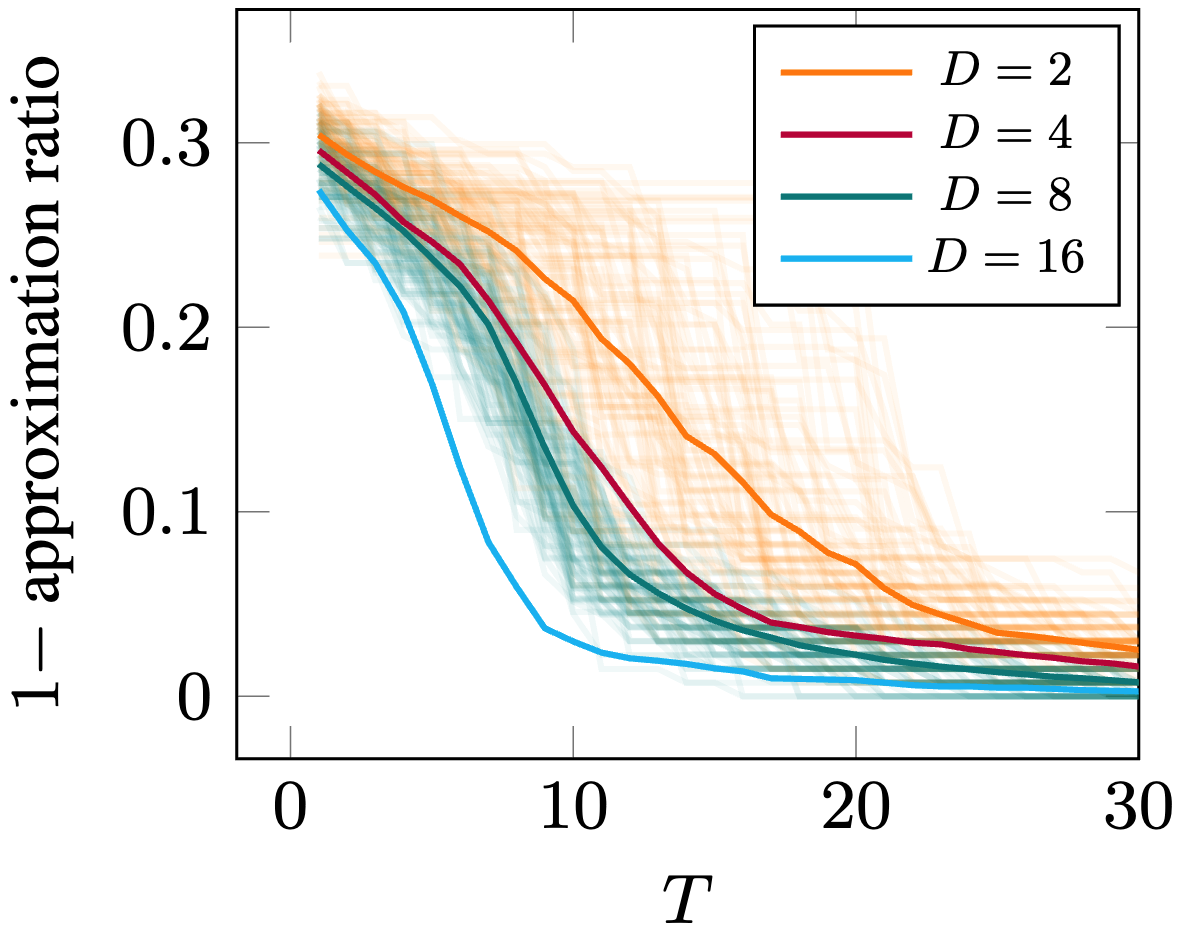}
\includegraphics[width=0.225\textwidth]{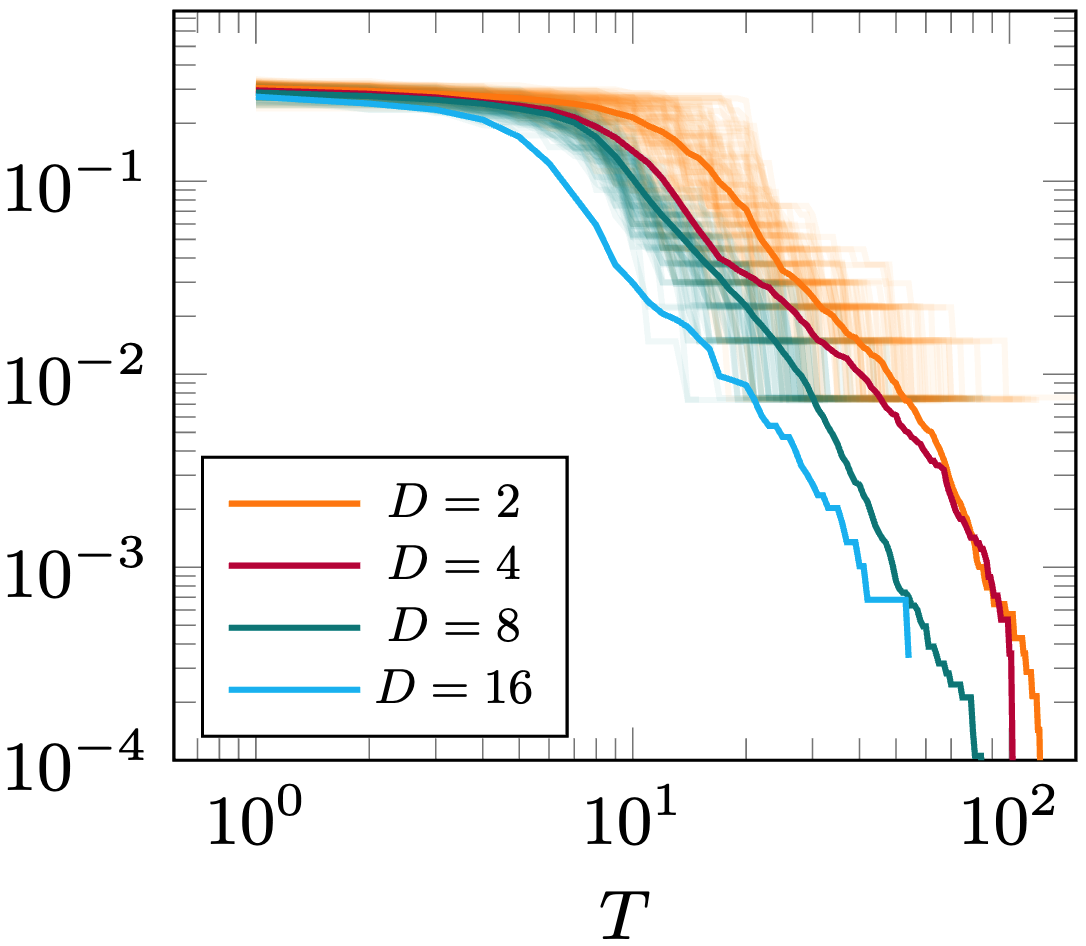}
\includegraphics[width=0.24\textwidth]{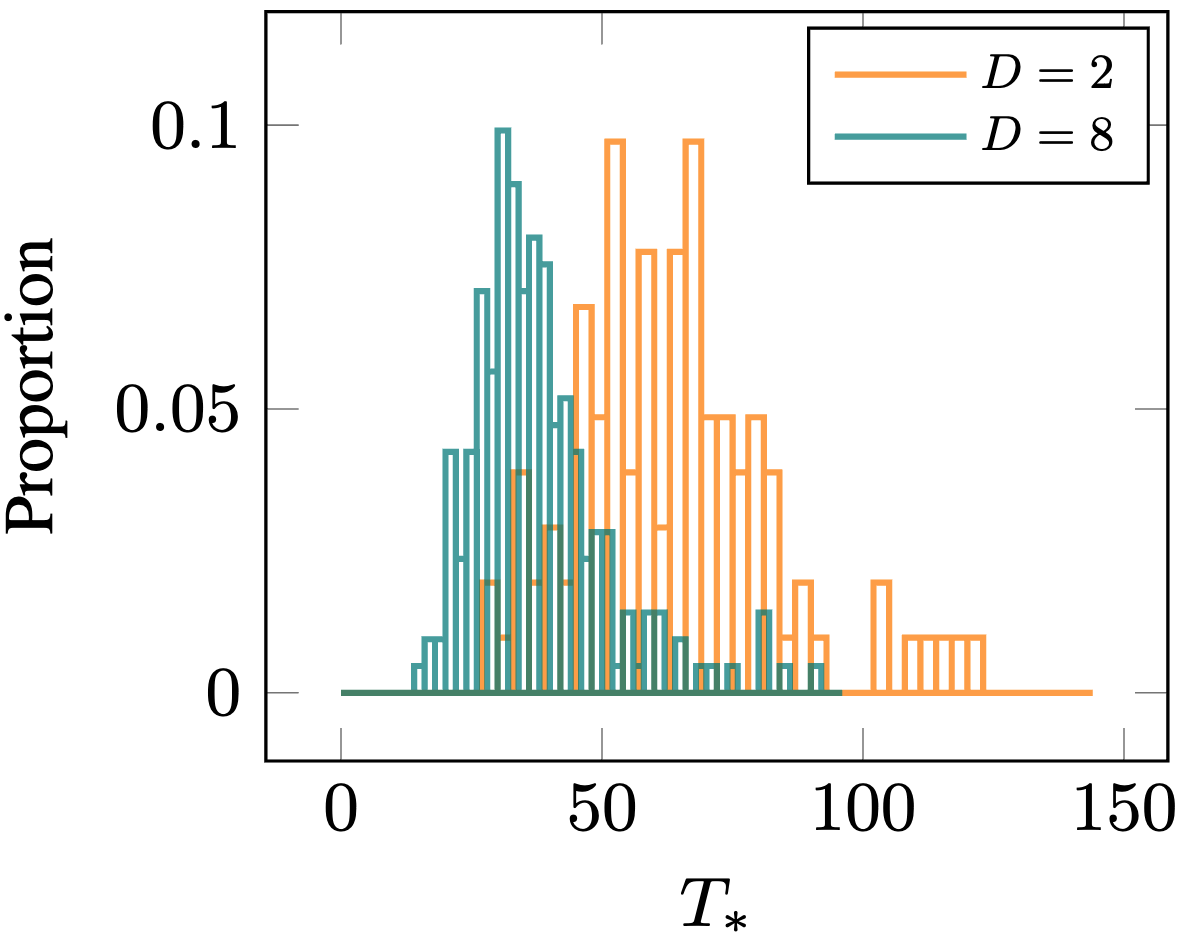}
\includegraphics[width=0.215\textwidth]{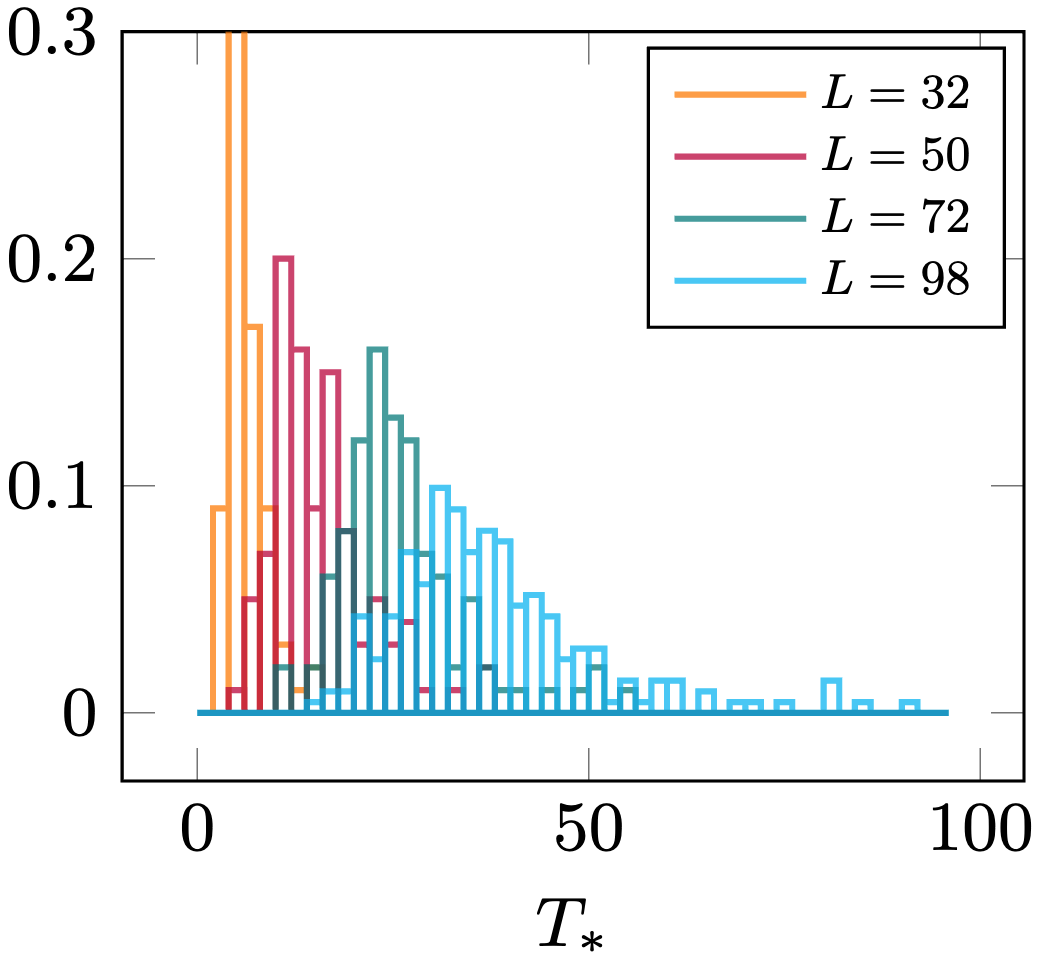}
\caption{\emph{Top:} One minus best approximation ratio found up to time $T$ with FAA, as a function of adiabatic time $T$, simulated with a MPS, on $3$-regular graphs of size $L=98$: individual samples (low opacity) and average (high opacity), for different bond dimensions $D$, in normal scale (left panel) and logarithmic scale (right panel), with Trotter step $\delta t=1/4$ and number of shots $N_S=10^3$. For better readability, individual samples are shown only for $D=2$ and $D=8$. 
\emph{Bottom:} Histograms of the minimal adiabatic times $T_*$ for which the optimal solution is obtained, for $L=98$ and different bond dimensions (left panel), and for $D=8$ and different system sizes (right panel).}
\label {trotter_fig2}
\end{center}
\end {figure}

\textbf{\emph{The Max-Cut problem.}}--- The \texttt{Max-Cut} problem is a paradigmatic classical optimization problem. Let $G$ be a graph on a set of $L$ vertices $V=\{1,...,L\}$ with a set of edges $E\subset V^2$. This problem consists in finding a partition $V=V_0\cup V_1$ with $V_0\cap V_1=\emptyset$ that maximises the number of edges $(i,j)\in E$ that connect the two subsets of vertices, namely $i\in V_0$ and $j\in V_1$. This number of edges is often referred to as the size of the cut. This problem is equivalently formulated as finding the ground state of the Ising Hamiltonian
\begin{equation}
    H=\sum_{(i,j)\in E}Z_iZ_j\,.
\end{equation}
Indeed, a basis of eigenstates of $H$ is trivially given by product states in the $Z$ basis $|b_1...b_L\rangle$ with $b_j\in\{0,1\}$, and their eigenvalue is $-c+(|E|-c)=-2c+|E|$ where $c$ is the number of edges $(i,j)\in E$ such that $b_i=0$ and $b_j=1$. Hence finding a product state that minimizes the energy is equivalent to finding a configuration that maximizes $c$, i.e. solving the \texttt{Max-Cut} problem on the graph. 
In this work we will focus on $3$-regular graphs, namely graphs where each vertex has exactly $3$ neighbours.

The \texttt{Max-Cut} problem for a generic graph is NP-hard \cite{lewis1983michael}. There is no known classical algorithm to solve \texttt{Max-Cut} for a generic graph whose runtime is polynomial in the number of vertices $L$, and it is strongly believed that such an algorithm does not exist. Finding a partition with a cut size $c'$ that approximates the maximal cut size $c$ up to a ratio $\frac{16}{17}\approx 0.941$, namely $c'>\frac{16c}{17}$, is also known to be NP-hard, and so cannot be solved in a time polynomial in $L$ neither \cite{haastad2001some}. In practice, there exist however highly optimized generic exact solvers that can solve graphs of intermediate sizes, despite their runtime being exponential in $L$ \cite{kugel2010improved}. For larger graphs or approximate solutions, there exists a plethora of approximate solvers that are used in the industry \cite{goemans1995improved,gurobi,DunningEtAl2018}. 
For $3$-regular graphs, there is a polynomial time algorithm with approximation ratio $\approx 0.9326$ \cite{halperin2004max}, and going beyond $\frac{331}{333}\approx 0.994$ is known to be NP-hard \cite{berman1999some}. 
Due to the high relevance of this kind of optimization problems in the industry, any improvement in the runtime of these solvers can have significant impact \cite{kochenberger2014unconstrained}.

\textbf{\emph{The adiabatic algorithm.}}--- One of the workhorse of quantum computing is the adiabatic algorithm and its variants. We consider the following time-dependent Hamiltonian for $0\leq s\leq 1$
\begin{equation}
    H(s)=-(1-s)\sum_{j=1}^L X_j+s\sum_{(i,j)\in E}Z_iZ_j\,.
\end{equation}
The ground state of $H(s)$ at $s=0$ is the product state $|++...+\rangle$, while the ground state of $H(s)$ at $s=1$ solves the \texttt{Max-Cut} problem on the graph. Let us consider $|\psi(t)\rangle$ a state equal to $|++...+\rangle$ at time $t=0$, and evolve it for a time $T$ with the time-dependent Hamiltonian $H(t/T)$, namely
\begin{equation}\label{adiabatic}
    \partial_t |\psi\rangle=iH(t/T)|\psi\rangle\,.
\end{equation}
No crossings can occur in $H(t/T)$ for $0<t<T$ due to the Perron-Frobenius theorem \cite{farhi2014quantum}. Hence, according to the adiabatic theorem, provided the time evolution is slow enough, i.e. if $T$ is large enough, then $|\psi(t)\rangle$ remains in the ground state of $H(t/T)$ for all $t$. In particular the state prepared at time $t=T$ becomes closer to a \texttt{Max-Cut} solution as $T$ grows larger. On a classical computer, computing this time evolution is strongly believed to be exponentially costly in $L$ or $T$ depending on the method used, see e.g. \cite{schuch2008entropy} for MPS simulations. On a quantum computer however, this time evolution can be computed with a runtime that is polynomial in system size $L$ and simulation time $T$ \cite{lloyd1996universal,suzuki1991general}. The scaling of the runtime with $L$ to solve \texttt{Max-Cut} with the adiabatic state preparation depends thus crucially on the simulation time $T(L)$ that is required for $|\psi(T)\rangle$ to have an appreciable overlap with a \texttt{Max-Cut} solution. 

Implementing the adiabatic evolution \eqref{adiabatic} on a gate-based quantum computer requires to discretize the time evolution, for example with a Trotter decomposition. We define the circuit
\begin{equation}
    U(M,\delta t)=W_{1}(\delta t)...W_{2/M}(\delta t)W_{1/M}(\delta t)\,,
\end{equation}
with $M$ an integer, $\delta t$ some time step, and where we denote for a real $0\leq s\leq 1$
\begin{equation}\label{ws}
    W_s(\delta t)=\prod_{(j,k)\in E}e^{is\delta t Z_jZ_k} \prod_{j=1,...,L} e^{-i(1-s)\delta t X_j}\,.
\end{equation}
Taking $\delta t=\frac{T}{M}$, we have according to the Trotter formula
\begin{equation}
    \underset{M\to\infty}{\lim} U(M,\tfrac{T}{M})|+...+\rangle=|\psi(T)\rangle\,.
\end{equation}
The number of steps $M$ to take to reach precision $\epsilon$ on the prepared state is $M\sim \frac{T^2L}{\epsilon}$ \cite{childs2021theory}. This gives a total gate count $\sim \frac{T^2L^2}{\epsilon}$. Previous works in different contexts have suggested more favorable scalings \cite{kovalsky2023self,yi2021success}. There exist other algorithms for running adiabatic evolution with a better scaling, but they are more involved \cite{berry2015simulating,berry2020time,granet2023continuous}. Implementations of the adiabatic algorithm for classical optimization problems have been done in the past in various works, whether classically simulated or on actual hardware \cite{farhi2001quantum,crosson2014different,hamerly2019experimental,zhou2020quantum,ebadi2022quantum,barends2016digitized,mehta2022quantum,vert2021benchmarking,farhi2012performance,ikeda2019application,denchev2016computational,otterbach2017unsupervised,albash2018demonstration,mohseni2022ising,santra2014max,king2022coherent,amaro2022filtering,keever2023towards}.

\textbf{\emph{The Floquet adiabatic algorithm.}}--- Instead of scaling the time step $\delta t$ to $0$, as would be done to minimize errors due to Trotterization, let us \emph{fix} the time step $\delta t$ to some finite value $\delta t=\frac{1}{n}$ for a real $n$, take the adiabatic time $T$ to be an integer multiple of $1/n$, and fix the number of Trotter steps $M=nT$. The operator that we implement is thus $\mathcal{U}(T)\equiv U(nT,\tfrac{1}{n})$ given by
\begin{equation}\label{mathcalu}
    \mathcal{U}(T)=W_{1}(\tfrac{1}{n})...W_{2/(nT)}(\tfrac{1}{n})W_{1/(nT)}(\tfrac{1}{n})\,,
\end{equation}
with $W_s(\delta t)$ defined in \eqref{ws}.

The operator $\mathcal{U}(T)$ \emph{does not} implement the standard adiabatic algorithm. Even when $T\to\infty$, the time step is finite and the time evolution alternates between purely $X_j$ evolution and purely $Z_jZ_k$ for a finite duration. In particular, the state of the system is never in the ground state of $H(s)$ when $0<s<1$, even when $T\to\infty$. However, the spectrum of the unitary operator $W_s(\delta t)$ when $s$ goes from $0$ to $1$, goes smoothly from $\{e^{-i\delta t x}\}$ with $x$ eigenvalues of $\sum_{j=1}^L X_j$, to $\{e^{i\delta t z}\}$ with $z$ the eigenvalues of $\sum_{(j,k)\in E}Z_jZ_k$. Let us write
\begin{equation}
    W_s(\delta t)=e^{i\delta tF_{\delta t}(s)}\,,
\end{equation}
with $F_{\delta t}(s)$ called Floquet Hamiltonian. When $T\to\infty$, because the Trotter step $\delta t$ is kept finite, we still alternate with time evolution $e^{is\delta t Z_jZ_k} $ and time evolution  $e^{-i(1-s)\delta t X_j}$. However, the rate at which $s$ is modified from one term $W_s(\delta t)$ to the next decreases with $T$. In the limit $T\to\infty$, we thus implement an adiabatic evolution with respect to $F_{\delta t}(s)$ instead of $H(s)$. Now, the crucial point is that, because $H(s=1)$ is a sum of commuting terms, we have
\begin{equation}
    F_{\delta t}(s=1)= H(s=1)\,.
\end{equation}
This holds true only when targeting classical optimization problems, where the Hamiltonian $H$ is classical. As a consequence, according to the adiabatic theorem, if there is no crossings in the Floquet Hamiltonian $F_{\delta t}(s)$, then a system initially prepared in the ground state of $\sum_{j=1}^L X_j$ will converge to the ground state of $H$ when $T\to\infty$.

\textbf{\emph{Floquet  adiabatic algorithm for \texttt{Max-Cut}.}}--- This observation motivates us to study the following heuristic algorithm that we implement for solving \texttt{Max-Cut}, dubbed Floquet  adiabatic algorithm (FAA) \cite{suzuki1990fractal,farhi2000quantum}. 

The algorithm takes as input a graph $G$ on $L$ vertices, and as parameters a time step $\delta t$ fixed, a number of shots $N_S$ and a stopping criterion. 

The algorithm proceeds as follows. We take an arbitrary initial value of adiabatic time $T$. We prepare the initial state in the product of $|+\rangle$ at each site,  apply $\mathcal{U}(T)$ defined in \eqref{mathcalu}, and then measure the state in the $Z$ basis, collecting $L$ measurement outcomes, one for each vertex. We repeat this $N_S$ times, obtaining $N_S$ collections of $L$ measurement outcomes. From them, we compute the value taken by $H=\sum_{(i,j)\in E}Z_iZ_j$ on each of the $N_S$ shots, and define $m_G(T)$ as the minimal value obtained from these $N_S$ shots. We repeat this process for the same graph $G$, but with increasing $T$, collecting a sequence $m_G(T_1),...,m_G(T_q)$, until we reach the stopping criterion decided in advance. Then $m_G^*$ is defined as the minimal value of the $m_G(T_i)$ over each of the times $T_i$ assessed. The configuration of spins $Z_i=\pm 1$ realizing this minimal energy can be read off from the corresponding measurement outcomes.

\textbf{\emph{Numerical simulations.}}--- The runtime of adiabatic algorithms is most often out of reach of analytical study. In order to evaluate the runtime of FAA, we now present numerical simulations. Exact simulation of FAA with state-vector simulations is exponentially costly in the number of vertices $L$, limiting this approach to $L\lessapprox 30$. To reach larger system sizes, we use Qiskit's MPS simulator with a finite bond dimension $D$ \cite{Qiskit}. This provides only an approximation to the exact FAA, but the approximation should improve as $D$ increases. The simulation runtime for one graph $G$ and one adiabatic time $T$ is then $\mathcal{O}(LTD^3)$. In the rest of this work, unless stated otherwise we will fix the number of shots per circuit to be $N_S=10^3$ and Trotter step $\delta t=1/4$.

\begin{table}[H]
\begin{center}
\begin{tabular}{|c||c|c|c|c|c|c|c|}
\hline
     $D$ & $2$ & $4$ & $8$ & $16$ & $32$& $64$ &$\infty$\\
    \hline
    \hline
      $L\leq 24$& $400$ & $400$ & $400$ & $1100$ & $1200$ &$450$  & $2625$\\
    \hline
      $24<L\leq 32$& $300$ & $300$ & $202$ & $500$ &$596$&$200$ & $118$\\
    \hline
      $32<L\leq 50$& $100$ & $100$ & $100$ & $662$ &$365$ &$301$ &\\
    \hline
      $50<L\leq 72$& $139$ & $102$ & $100$ & $270$ &$49$ & $95$&\\
    \hline
      $L= 98$& $104$ & $100$ & $212$ & $22$ & & &\\
    \hline
      $L= 128$& $103$ &  & $41$ &  & & &\\
    \hline
      $L= 162$& $114$ &  &  &  & & &\\
    \hline
      $L= 200$& $93$ &  &  &  & & &\\
    \hline
      $L= 242$& $44$ &  &  &  & & &\\
    \hline
      $L= 288$& $19$ &  &  &  & & &\\
    \hline

\end{tabular}
\end{center}
\caption{Number of instances of $3$-regular graphs on which FAA with $\delta t=1/4$ has been run, for different system sizes $L$ (rows) and different bond dimensions $D$ (columns). The optimal solution computed with \texttt{akmaxsat} has always been found with FAA provided $T$ is taken large enough.}
\label {tabular}
\end{table}

We start with presenting in Figure \ref{trotter_fig2} some runs of the algorithm in the following setting. We fix the number of vertices to $L=98$ and we generate random $3$-regular graph $G$ on these $L$ vertices. Using the exact \texttt{akmaxsat} solver \cite{kugel2010improved}, the maximal cut on these graphs can be found classically in reasonable time on a laptop for system sizes up to $L\sim 300$, and we denote $m_G$ the exact minimal energy of graph $G$.  For each of these graphs, we obtain some numerical result $m_G(T')$ for $T'=1,2,...,T$, and define the minimum $m^*_G(T)=\min_{1\leq T'\leq T}m_G(T')$. From this, one obtains an approximation ratio $r(T)$ for the maximal cut in this graph, as $r(T)=\frac{-m^*_G(T)+3L/2}{-m_G+3L/2}$. Then we report in the top panels of Figure \ref{trotter_fig2} the value $1-r(T)$ as a function of $T$, plotting both individual trajectories for each random graph $G$, and their average. In the bottom panels, we report histograms of the minimal adiabatic time $T_*(G)$ for which the optimal solution is found, across different graphs $G$.

\begin{figure*}[tb]
\begin{center}

\includegraphics[width=0.225\textwidth]{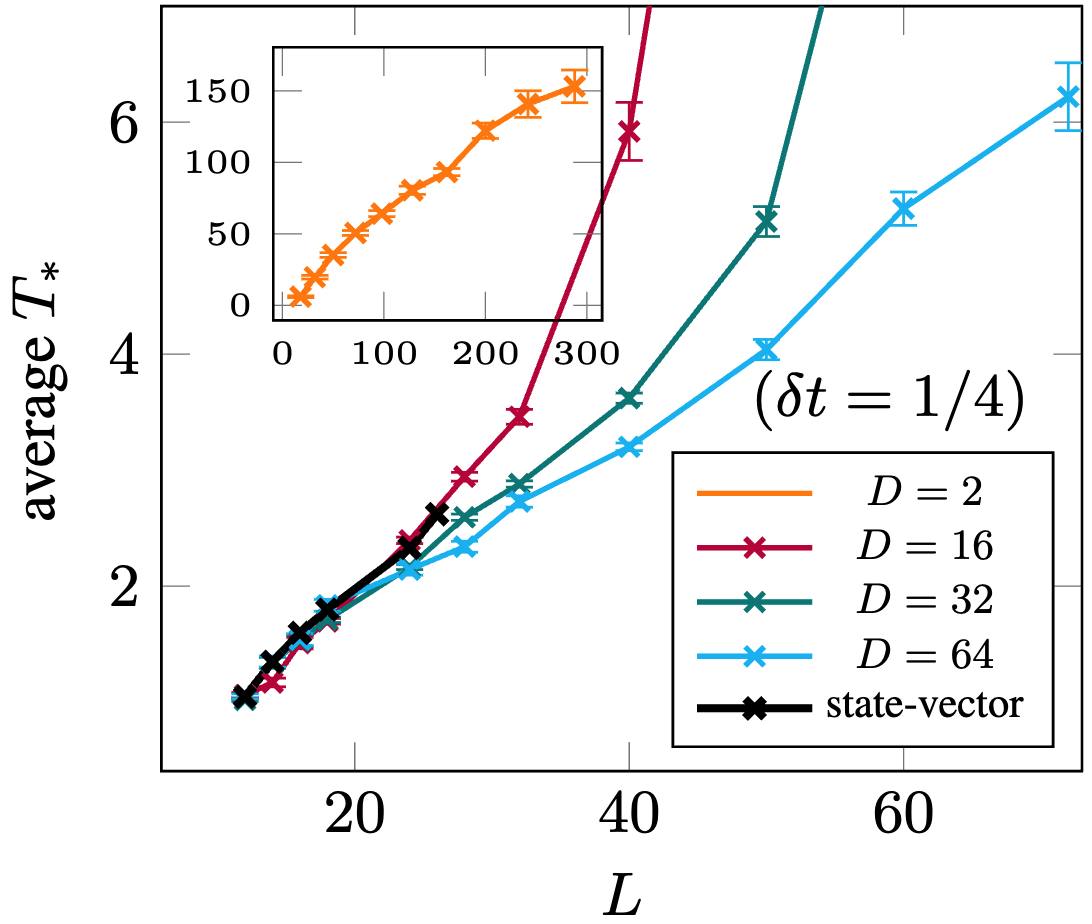}
\includegraphics[width=0.225\textwidth]{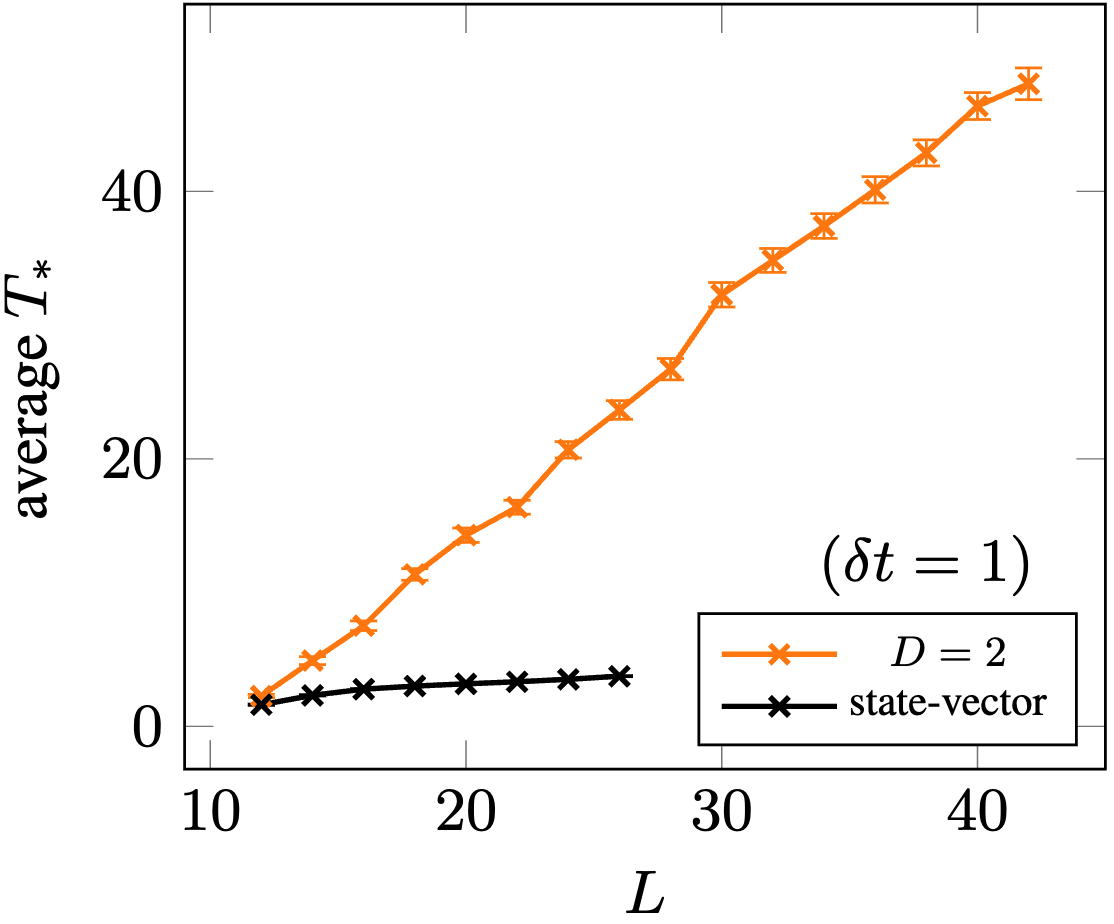}
\includegraphics[width=0.225\textwidth]{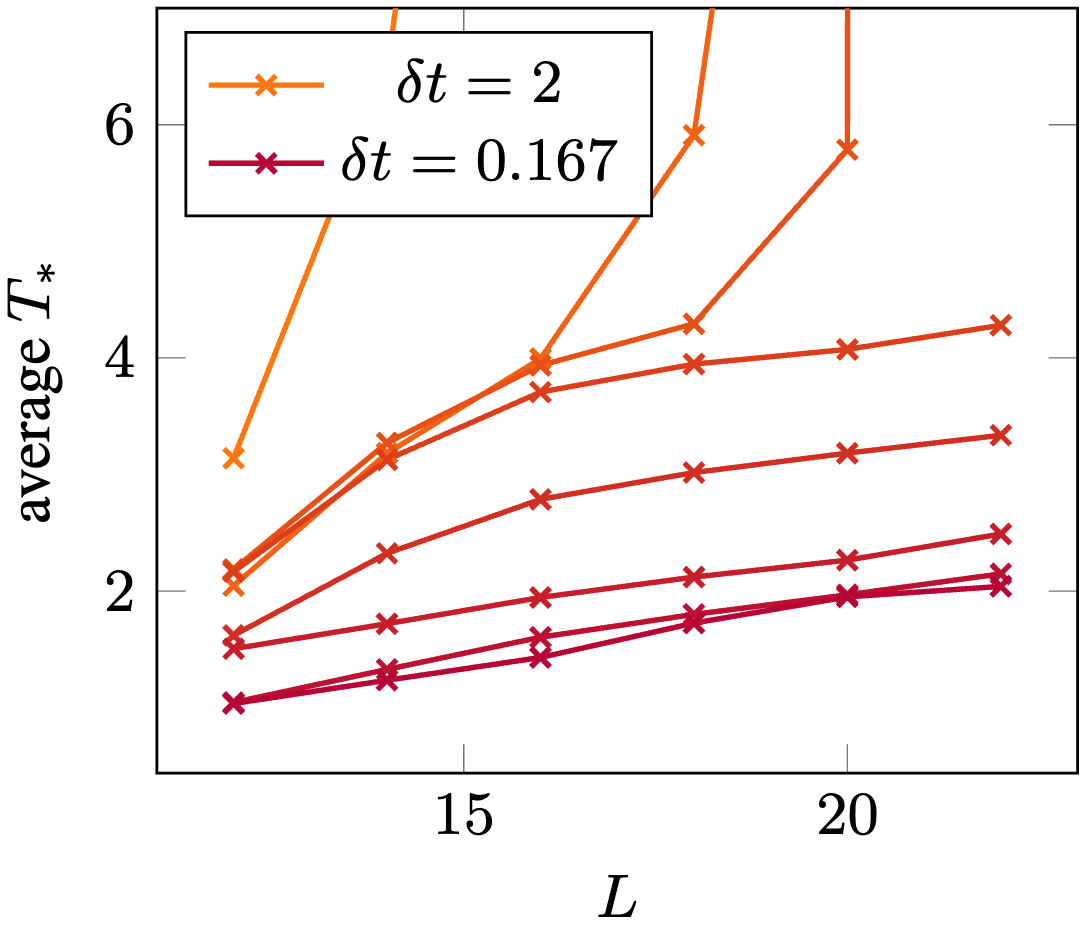}
\includegraphics[width=0.225\textwidth]{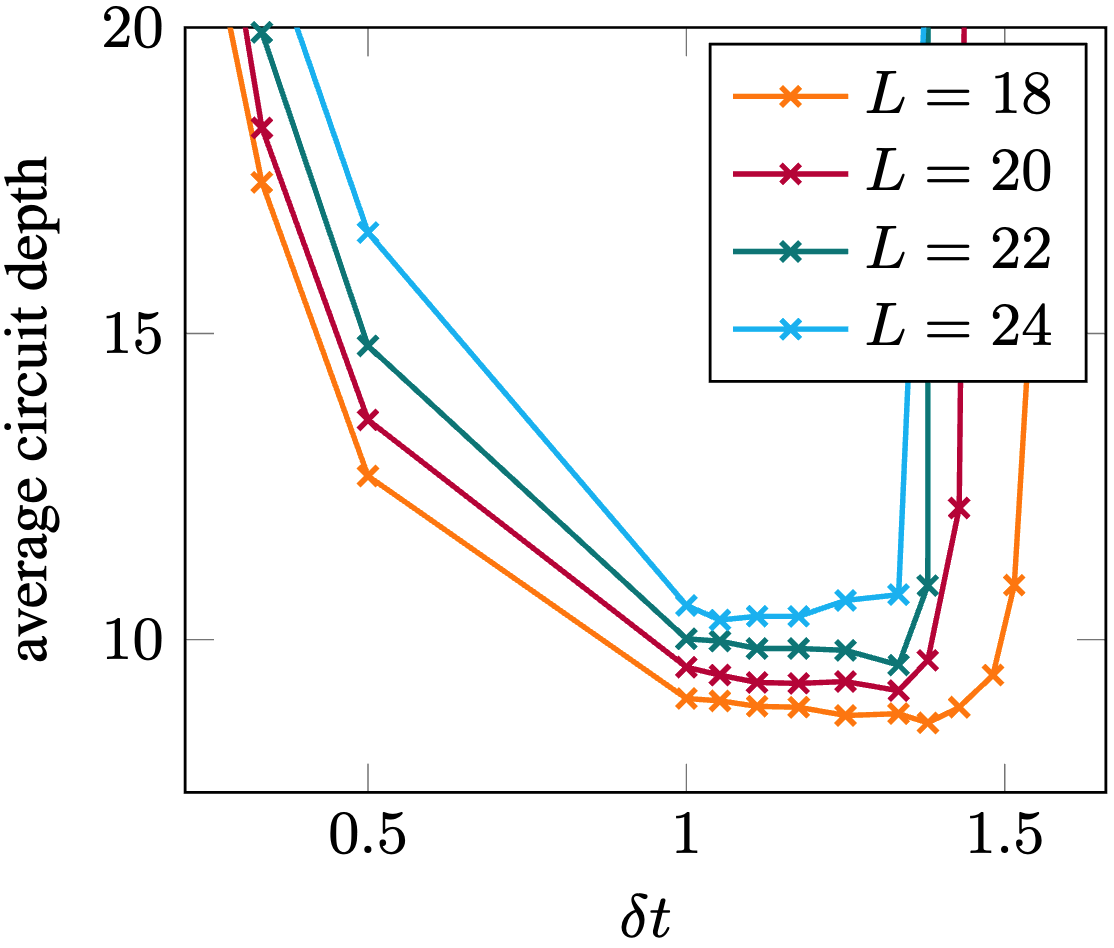}
\caption{ From left to right: \emph{First panel:} Average minimal adiabatic time $T_*$ required to find the optimal solution in random $3$-regular graphs $G$, as a function of $L$, using MPS simulations with different bond dimensions $D$, and using state-vector simulations (black), for $N_S=10^3$ and $\delta t=1/4$ (inset shows $D=2$). \emph{Second panel:} Same as first panel, but with $\delta t=1$. \emph{Third panel:} Same as first panel,  but for different values of $\delta t=\frac{1}{n}$ with $n=0.5$, $0.66$, $0.7$, $0.75$, $1$, $2$,  $4$, $6$ (from orange to purple, or from top to bottom), using only state-vector simulations. \emph{Fourth panel:} Average circuit depth $3T_*/\delta t$, as a function of Trotter step $\delta t=\frac{1}{n}$, for different system sizes, using state-vector simulations.
}
\label {trotter_fig3}
\end{center}
\end {figure*}

From these results, we make the following three observations. Firstly and most importantly, for all the graphs we tried, we \emph{always} find the optimal solution $m_G$ provided the adiabatic time $T$ is large enough. This holds true despite the total number of configurations sampled per graph $N_ST\lessapprox 10^5$ being well below the total number of configurations $2^L\sim 10^{30}$. We emphasize that, remarkably, this holds true \emph{even} for bond dimension $D=2$. In Table \ref{tabular} we indicate the number of instances tested in this work for $\delta t=1/4$ across different system sizes $L$ and bond dimensions $D$, giving a statistical upper bound for the proportion of graphs for which FAA would fail for different $D$ and $L$. Second, we observe that while variations of the curve $m_G(T)$ are quite pronounced for $D=2$ across different graphs $G$, they are much smaller for bond dimension $D=8$. This suggests that the exact time evolution (obtained for $D\to\infty$) would present only low variation for different graphs $G$. This is confirmed by the bottom left panel, where the distribution of the smallest adiabatic times required to find the optimal solution gets narrower as the bond dimension is increased. On the right bottom panel, we see conversely that at fixed bond dimension, this distribution becomes broader as the system size is increased, reflecting the fact that a fixed bond dimension $D$ will capture a smaller portion of the Hilbert space as $L$ increases. Third, the logarithmic scale plot clearly displays two regimes. There is a first regime up to e.g. time $T\approx 7$ for $D=8$ where the improvement is slow. After that point, there is a sharp transition to a fast decay of the energy error, compatible with a power-law behaviour $\sim 1/T^2$. There is possibly a third regime with exponential decay at larger times. These scalings would matter when targetting a certain approximation ratio $<1$ instead of the optimal solution.

We move on to the study in Figure \ref{trotter_fig3} of the scaling with number of vertices $L$ of the minimal adiabatic time $T_*(L)$ to find the optimal solution. Indeed, a crucial point in assessing the viability of classical or quantum algorithm is to understand how its complexity scales with input parameters. Given a $3$-regular graph $G$ on $L$ vertices, we run FAA starting from adiabatic time $T=1$ and increasing $T$ by $1$ at each iteration, until the maximal cut has been found (computed with \texttt{akmaxsat}), denoting $T_*$ the corresponding value of $T$. The first conclusion from this numerics is that, again, we \emph{always} end up finding the optimal cut when increasing $T$, even if the fraction of configurations explored is several tens of orders of magnitude below $1$. In the leftmost panel of Figure \ref{trotter_fig3}, we show $T_*(L)$ the average of $T_*$, as a function of number of vertices $L$. Up to system sizes studied, the numerics could be compatible with a linear scaling for bond dimension $D=2$ or $D=64$ for example, although the scaling is strongly expected to be exponential in $L$ due to NP-hardness. This exponential behaviour should become visible at larger values of $L$, but these sizes become unreachable by exact solvers. In the second panel of Figure \ref{trotter_fig3}, we plot the same numerics but for Trotter step $\delta t=1$. The linear scaling for these small to intermediate system sizes is here well visible, as well as the considerably smaller average $T^*$ obtained with state-vector simulation (or $D=\infty$) compared to MPS with $D=2$.

Next, we investigate in the two rightmost panels of Figure \ref{trotter_fig3} the effect of the finite Trotter step $\delta t=\frac{1}{n}$. For $\delta t=\pi/2$, since we have $e^{i\delta t H(s)}=\pm e^{i\frac{\pi s}{2}(\sum_j X_j+\sum_{(j,k\in E)}Z_jZ_k)}$, FAA cannot succeed as the time evolution is not that of an adiabatic evolution. We thus expect FAA to fail at least around and above $\delta t \sim \pi/2$. In practice, using state-vector simulations up to $L=22$, we observe in the third panel of Figure \ref{trotter_fig3} that for $n\gtrapprox 0.8$ the growth of $T_*(L)$ with $L$ is mild, and we always find the optimal solution with an adiabatic time $T$ small or moderate. However for $n\lessapprox 0.8$ the time $T_*$ blows up with $L$, and this increase is mainly due to outlier graphs for which FAA fails, i.e. require a very large $T$ to find the solution by just random sampling (with our protocol where we try every integer value of $T$, we typically expect to find the optimal solution for $T_*\lessapprox\frac{2^L}{N_S}$).  In the fourth panel, we then show the circuit depth $3nT_*$ as a function of $\delta t=1/n$. We observe that the curve reaches a minimum between $\delta t=1$ and $\delta t=1.3$, indicating the optimal range of Trotter steps on which to run FAA. This optimal value is very far from $\delta t=0$ required to implement the exact (non-Floquet) adiabatic evolution. 

\begin{figure}[tb]
\begin{center}

\includegraphics[width=0.225\textwidth]{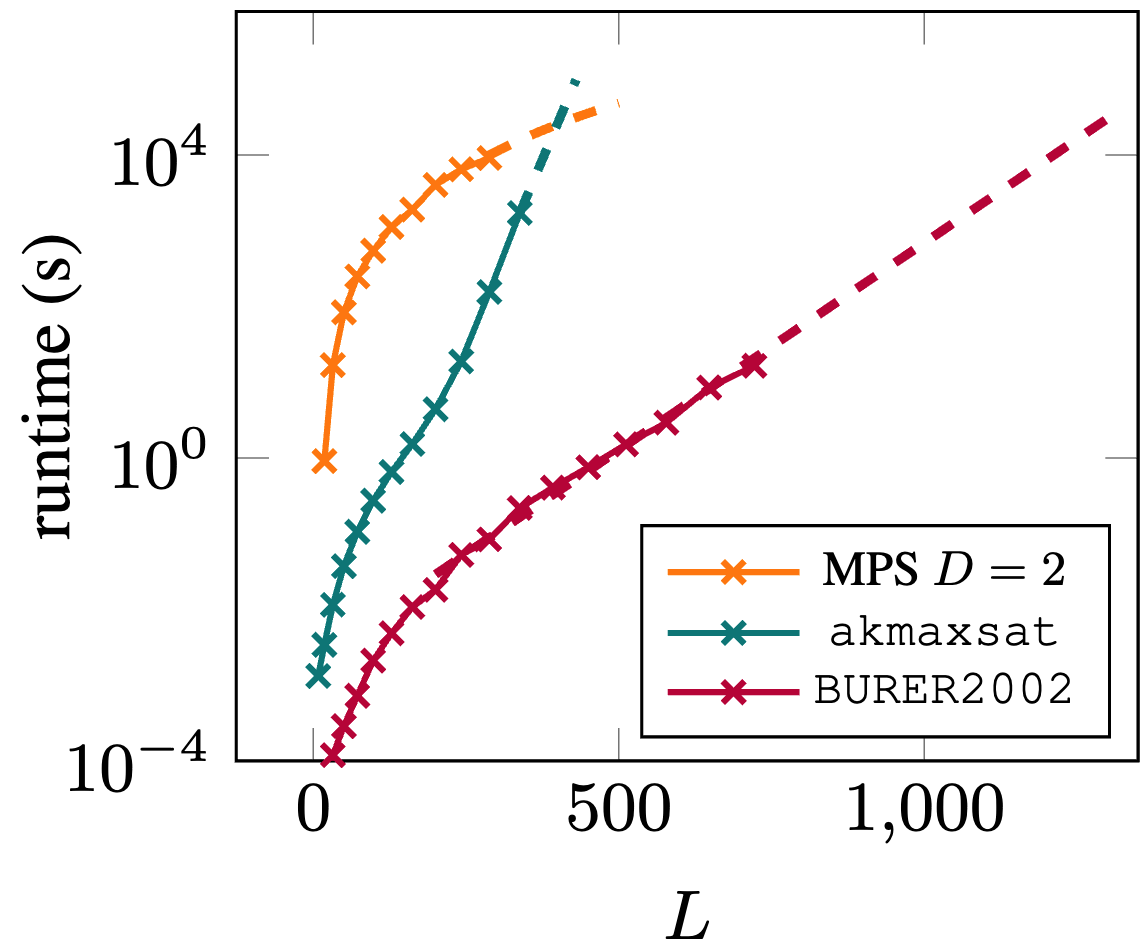}
\includegraphics[width=0.225\textwidth]{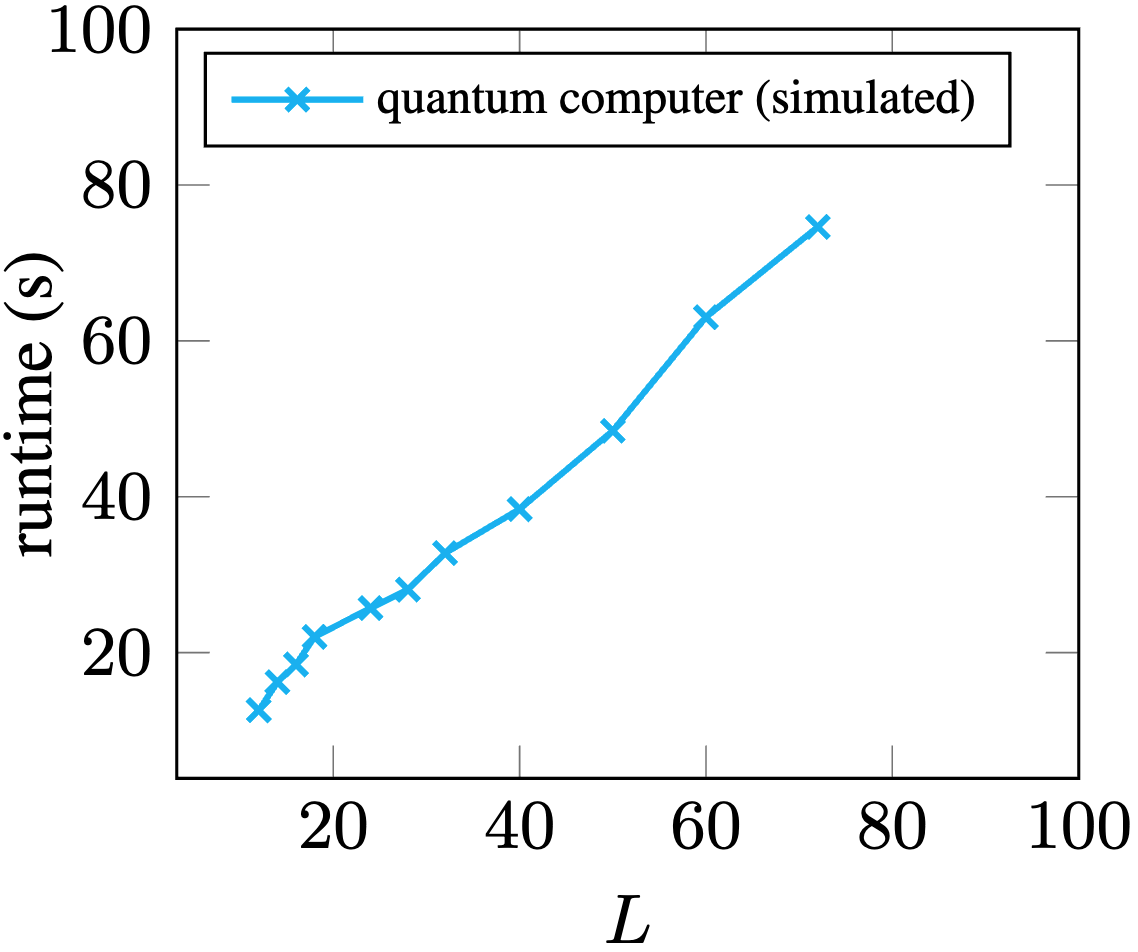}

\caption{ \emph{Left panel:} Average runtime in seconds to find the optimal solution on a random $3$-regular graph, as a function of $L$, with the \texttt{akmaxsat} exact solver, with the \texttt{BURER2002} approximate solver, and with MPS simulations with $D=2$ for Trotter step $\delta t=1/4$. Teal and orange are evaluated on a Intel x86/64 processor, with orange evaluated only at $L=50$, and then using the scaling $LT_*(L)^2$. Dashed lines indicate extrapolation, with the \texttt{BURER2002} runtime fitted with $\exp(-6.06+0.0126L)$. The curve of \texttt{BURER2002} for $L\geq 300$ is only a lower bound of the actual runtime, due to absence of guarantee of convergence. \emph{Right panel:} Estimated average runtime in seconds to find the optimal solution on a random $3$-regular graph by implementing FAA on a quantum computer, with assumptions detailed in the text.
}
\label {trotter_fig4}
\end{center}
\end {figure}

In Figure \ref{trotter_fig4} we then plot as a function of system size the runtime of FAA, that of the exact classical solver \texttt{akmaxsat} \cite{kugel2010improved}, and that of the approximate classical solver \texttt{BURER2002} from the MQLib library \cite{DunningEtAl2018}. They were both used for benchmarking the \texttt{Max-Cut} problem in previous works \cite{santra2014max,guerreschi2019qaoa,lykov2023sampling}. In the left panel, we show the classical simulation of FAA with MPS which scales as $D^3LT_*(L)^2$. Although $T_*(L)$ decreases with bond dimension $D$, due to the $D^3$ scaling of MPS simulations, we find that $D=2$ provides the best runtime. We see that FAA simulated with a MPS is expected to perform better than \texttt{akmaxsat} around system size $L=400$.

\textbf{\emph{On a quantum computer.}}--- We now describe the performance of running FAA on a quantum computer. We assume a runtime of $1$ms per two-qubit gate, and possible parallelization of $L/2$ gates in the quantum circuit since gates that apply on disjoint edges can be performed simultaneously, in contrast with classical simulations. As a consequence the runtime of the quantum implementation of FAA only scales as $T_*(L)^2$, and the minimal adiabatic time $T_*(L)$ corresponds to that of MPS with $D=\infty$. In the right panel of Figure \ref{trotter_fig4}, we show  an estimation of this runtime for a single adiabatic time $T=T_*$ (i.e., without doing the step by step increase in $T$), with $T_*(L)$ obtained with classical MPS simulation with  $D=64$. For these system sizes $L\leq 72$, the runtime as a function of $L$ is seen to be compatible with a linear behaviour. This linear behaviour is expected to eventually disappear at larger $L$, but it could in principle hold for a certain interesting and useful range of intermediate sizes. From the left panel, we see that system sizes $L=2000$ become completely out of reach of classical solvers, whether exact or approximate. Assuming the linear behaviour holds up to these sizes, and extrapolating our numerics for $T_*$ with $\delta t=1/4$ up to $L\sim 10^3$, we find $T_*\sim 10^2$, so around $12LT_*\sim 10^6$ two-qubit gates. Taking into account a few trials at different values of $T$, this would take around one day runtime. 
 
 This favorable extrapolated runtime is only possible due to keeping finite Trotter steps in FAA, in which the circuit depth is only $\mathcal{O}(nT_*)$. In contrast, the implementation of quantum (non-Floquet) adiabatic algorithms require a circuit depth $\mathcal{O}(T_*^2 L/\epsilon)$. Moreover we saw in the two rightmost panels of Figure \ref{trotter_fig3} that the adiabatic time required when $\delta t\to 0$ is only slightly smaller than for e.g. $\delta=1/4$. Even neglecting the factor $1/\epsilon$, this represents an improvement $\mathcal{O}(T_*L)$ in circuit depth and number of gates. For sizes $L\sim 10^3$ beyond what can be solved with classical solvers, this represents a saving of order $\sim 10^5$ both in circuit depth and gate count. 

 Let us finally comment on the effect of hardware noise, which is the main limiting factor of current and near-term quantum processors. Any source of noise that can be described by a quantum channel can be seen as applying Kraus operators with some probability $p$ after each gate in the circuit \cite{nielsen2010quantum}.
Then, in a circuit with $N$ noisy gates, there will always be a fraction $(1-p)^N$ of all the shots that are effectively noiseless. In general quantum computing applications, mitigating or correcting the noise requires to know in which shots an error occurred. However, in case of a classical optimization problem in NP, given a list of measurement outcomes in the $Z$ basis, one can compute the energy and check whether it is lower than previous measurement outcomes, without knowing whether an error occurred or not. Thus, one can obtain an estimate of the minimal energy from low-fidelity simulations at the only cost of extra shots (albeit exponentially many in $pN$). Compared to applications that require a global fidelity of order $\sim 1$ for noise to be mitigated, this entails a significant resource saving, as one can save a factor $10$ on the gate error at the cost of doing $\sim 10^4$ more shots. We note that this argument is not specific to FAA. In the case of the \texttt{Max-Cut} problem on a $3$-regular graphs, our estimate was that a circuit with around $\sim 10^6$ two-qubit gates was necessary to compete with classical computers, whence a required gate error probability $p\sim 10^{-5}$. This would set the resources to $\sim 10^{-5}$ gate error probability and $\sim 10^3$ qubits for a quantum computer to compete with classical solvers at the specific task of finding the optimal cut in a $3$-regular graph. Of course, these numbers rely on extrapolating trends observed numerically for system sizes $L\lessapprox 100$ that could break at larger $L$. However, they are encouraging and suggest a direction to explore further for industrial applications in the intermediate term. Before that, the classical verifiability of the algorithm presented here can be used as to benchmark quantum computers on application-oriented problems in large variety of circuit depths and number of qubits.

We thank David Amaro and Frederic Sauvage for useful comments on the manuscript. We thank Fermioniq for their MPS simulator initially used in this project \cite{fermioniq}. E.G. acknowledges support by the Bavarian Ministry of Economic Affairs, Regional Development and Energy (StMWi) under project Bench-QC (DIK0425/01).

%


\end{document}